\pdfoutput=1
\documentclass[preprint]{ptephy_v1}

\preprintnumber{YITP-19-16}

\usepackage{amsmath,amsthm,amssymb,amsfonts}
\usepackage{color}

\begin{document}

\title{Meson deformation by magnetic fields in lattice QCD}

\author{Koichi~Hattori$^{a}$}
\author{Arata~Yamamoto$^{b}$}

\affil{$^{a}$Yukawa Institute for Theoretical Physics, Kyoto University, Kyoto 606-8502, Japan}
\affil{$^{b}$Department of Physics, The University of Tokyo, Tokyo 113-0033, Japan}

\begin{abstract}
We study light meson properties in a magnetic field, 
focusing on a charged pion and a charged and polarized rho meson, in quenched lattice QCD.
The gauge-invariant density-density correlators are calculated
to investigate the deformation caused by the magnetic field. 
We find that these mesons acquire elongated shapes along the magnetic field.
The magnitude of the deformation is about 10-20 \% when the strength of the magnetic field 
is of the order of the squared unphysical pion mass. 
\end{abstract}

\subjectindex{B64}

\maketitle

\section{Introduction}

Strong magnetic fields have been implemented in lattice QCD simulation in the last decade. 
The large field strength, even beyond the QCD scale, provides us with 
opportunities to study interplay between QED and QCD, of which the typical scales are, otherwise, 
separated as known as the hierarchy of fundamental forces. 
Phenomenologically, the relativistic heavy-ion collisions and neutron stars/magnetars are thought to be 
accompanied by strong magnetic fields (see Refs.~\cite{Harding:2006qn,Hattori:2016emy} and references therein). 
The first-principle calculation by lattice QCD and phenomenological studies in those systems 
have been driving each other to deepen our understanding of intriguing phenomena in strong magnetic fields, 
such as the chiral magnetic effect \cite{Buividovich:2009wi,Yamamoto:2011gk,Yamamoto:2011ks,Bali:2014vja}. 

Lattice QCD simulation is particularly useful to investigate low-energy QCD in magnetic fields, 
while dynamics of the quark-gluon plasma can be also investigated with analytic methods, 
e.g., perturbative transport theory in magnetic fields \cite{Fukushima:2015wck, Hattori:2016cnt, 
Hattori:2016lqx, Fukushima:2017lvb, Li:2017tgi, Hattori:2017qih}. 
For a prominent example, lattice QCD simulation discovered the ``inverse magnetic catalysis'' phenomenon 
which indicates a decreasing behavior of the chiral transition temperature 
when the magnetic field strength is increased \cite{Bali:2012zg,Bruckmann:2013oba} (see also Refs.~\cite{DElia:2011koc,Ilgenfritz:2013ara,DElia:2018xwo}). 
It was a surprising discovery since the ``magnetic catalysis'' phenomenon, the chiral symmetry breaking 
induced by an arbitrarily weak attractive interaction, had been known 
on the basis of an analogy of the transition to superconducting phase at a finite density. 
The heart of the analogy was clearly identified with the dimensional reduction occurring 
in the low-energy excitations near the Fermi surface and in the lowest Landau level \cite{Gusynin:1994xp}. 
Therefore, as known in superconductivity, the phase transition temperature is expected to be the same order 
in magnitude as the dynamically generated gap at zero temperature which grows 
as the magnetic field strength is increased \cite{Gusynin:1997kj, Lee:1997uh,Fukushima:2012xw}. 
In contrast to this weak-coupling scenario, 
the inverse magnetic catalysis observed in the lattice QCD simulation indicates 
the opposite dependence of the transition temperature on the magnetic field strength. 
This qualitative discrepancy exemplifies the necessity for understanding the interplay between magnetic fields and nonperturbative gluodynamics. 

Lattice QCD simulation is a good tool to understand such interplay.
Since the fundamental degrees of freedom in low-energy QCD are hadrons as a consequence of color confinement,
the effects of magnetic fields result in the modification of hadron spectroscopy and hadron structure \cite{Lee:2008qf,Bali:2011qj,Hidaka:2012mz,Primer:2013pva,Beane:2014ora,Chang:2015qxa,Agadjanov:2016cjc,Bali:2017ian,Bali:2018sey}. 
In this work, we investigate the structures of light mesons with lattice QCD simulation. 
More specifically, we discuss the deformation of their shapes in distinct magnitudes of magnetic fields. 
We analyze it by using the gauge-invariant density-density correlator \cite{Burkardt:1994pw,Alexandrou:2002nn,Alexandrou:2003qt}. 

Here, it is worth mentioning an interesting idea that the QCD thermodynamics including the inverse magnetic catalysis
may be understood with the hadronic degrees of freedom, i.e., the hadron resonance gas model 
in magnetic fields \cite{Endrodi:2013cs, Hattori:2015aki, Fukushima:2016vix}. 
However, a naive hadronic model misses the internal structure or compositeness of hadrons, 
and the obtained results sometimes deviate from the principles in QCD. 
The lattice QCD simulation on hadron properties provides useful information 
for constructing the equation of state consistent with QCD 
as well as for understanding their one-body properties in its own 
(see Refs.~\cite{Fukushima:2012kc, Kojo:2012js, Hattori:2015aki} for 
construction of quark models in magnetic fields). 
We also note that the quarkonium spectroscopy and deformation of the wave functions 
were investigated with the Cornell potential model \cite{Alford:2013jva, Suzuki:2016kcs, Yoshida:2016xgm}. 
Studies on the heavy and light sectors are complementary to each other. 

The rest of this paper is organized as follows. 
In the next two sections, we first provide the lattice QCD formalism
to measure the deformation of mesons in magnetic fields, and then show the numerical results. 
We discuss possible origins of the deformation and give our perspective in Sec.~\ref{sec:perspective}. 
Some numerical check for the simulation is given in an appendix.

\section{Formalism}

We explain how to introduce external U(1) magnetic fields to lattice QCD and how to measure the spatial distribution of hadrons in lattice QCD.
The lattice size is denoted by $N_s^3\times N_t$ and the lattice spacing is denoted by $a$.
Periodic spatial boundary conditions are assumed.

In two-flavor lattice QCD, the quark action is given by
\begin{equation}
 S_f = \sum_{\rm site} \left( \bar{u} D[U_\mu] u + \bar{d} D[U_\mu] d \right)
,
\end{equation}
where $D[U_\mu]$ is the lattice Dirac operator and a functional of the SU(3) link variable $U_\mu$.
To introduce external electromagnetic fields, the Dirac operator is replaced as $D[U_\mu] \to D[V_\mu U_\mu]$ with the U(1) link variable $V_\mu$.
Let us consider a homogeneous magnetic field in the $z$ direction.
The U(1) link variable is set to be \cite{AlHashimi:2008hr}
\begin{equation}
\begin{split}
\label{eqV}
 V_1 &= \exp (-iqBN_say) \quad {\rm at} \ x=N_sa,
\\
 V_2 &= \exp (iqBax),
\\
 V_\mu &= 1 \quad {\rm for} \ {\rm other} \ {\rm components}.
\end{split}
\end{equation}
The magnetic field is quantized as $qB = 2\pi n/N_s^2$ $(n\in \mathbb{Z})$ due to periodic boundary conditions.
The electric charge $q$ is $2e/3$ for $u$ quarks and $-e/3$ for $d$ quarks.
The resultant quark action is
\begin{equation}
\label{eqSf}
 S_f = \sum_{\rm site} \left( \bar{u} D[V_\mu^{(u)} U_\mu] u + \bar{d} D[V_\mu^{(d)} U_\mu] d \right)
.
\end{equation}
The Dirac operator is flavor-dependent because of the difference of electric charges.
While the explicit form of the U(1) link variable \eqref{eqV} assumes a peculiar gauge choice, the action \eqref{eqSf} is gauge invariant.

\begin{figure}[t]
\begin{center}
 \includegraphics[width=.6\textwidth]{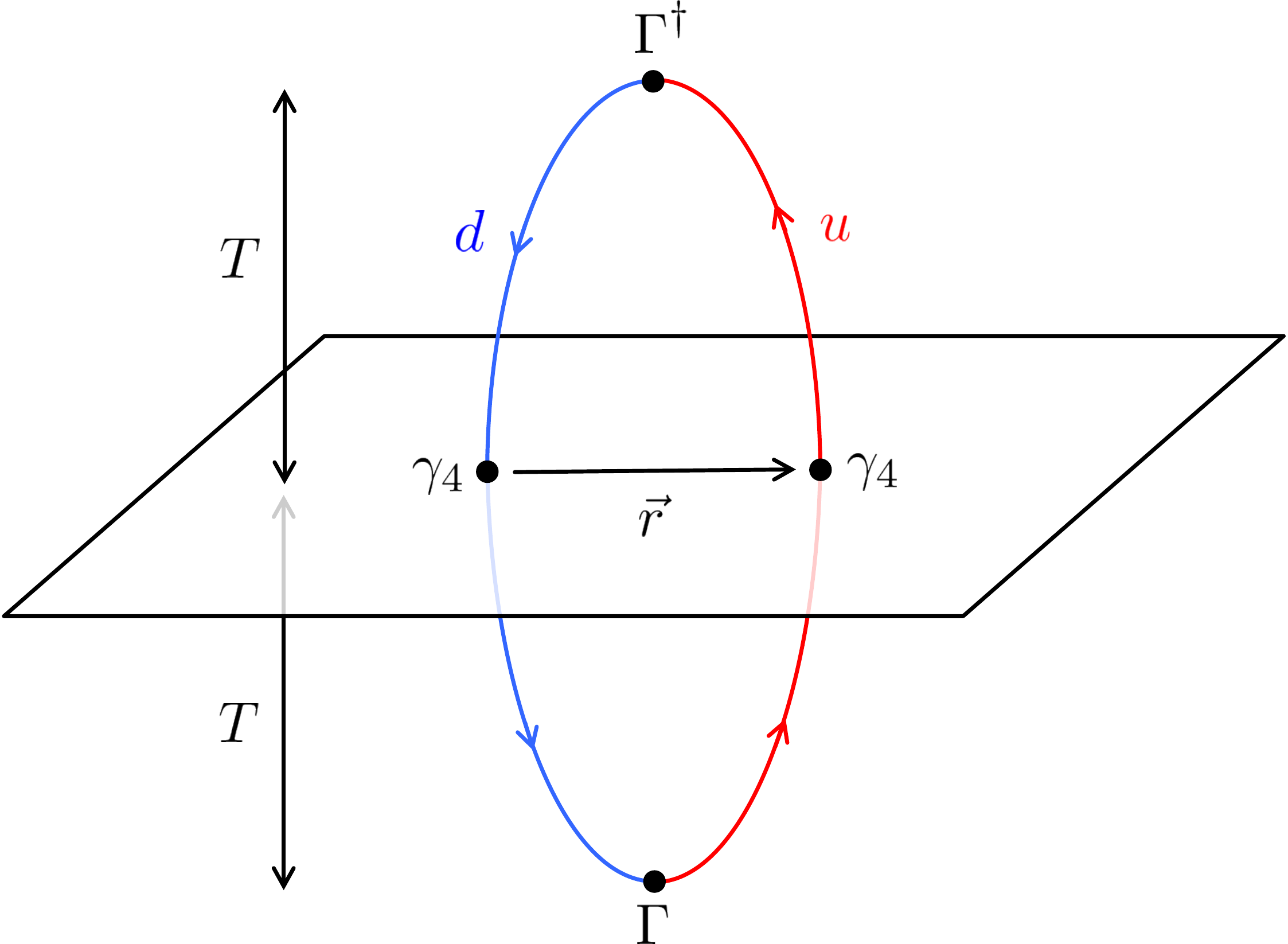}
\caption{
\label{fig}
The density-density correlators $C_\pi$ and $C_\rho$.
The red line is the $u$-quark propagator and the blue line is the $d$-quark propagator.
The matrix $\Gamma$ stands for $\gamma_5$ in $C_\pi$ and $(\gamma_1-i\gamma_2)/\sqrt{2}$ in $C_\rho$.
}
\end{center}
\end{figure}

We analyze the positively charged pion
\begin{equation}
\label{eqpi}
 \pi \equiv \bar{d} \gamma_5 u
\end{equation}
and the positively charged and polarized $\rho$-meson
\begin{equation}
\label{eqrho}
 \rho \equiv \bar{d} \left( \frac{\gamma_1-i\gamma_2}{\sqrt2} \right)u
.
\end{equation}
This polarization mode has the lightest mass spectrum among the $\rho$-meson states in magnetic fields 
as expected from the Zeeman splitting within the hadronic picture. 
There are several ways to measure the spatial distribution of hadrons in lattice QCD.
We adopt the density-density correlators in hadrons~\cite{Burkardt:1994pw,Alexandrou:2002nn,Alexandrou:2003qt}.
The correlators are defined by
\begin{eqnarray}
 C_\pi(\vec{r})  &=& \frac1{N_s^3} \sum_{\vec R} \langle \pi | n_u(\vec{R}+\vec{r}) n_d(\vec{R}) | \pi \rangle 
\\
 C_\rho(\vec{r}) &=& \frac1{N_s^3} \sum_{\vec R} \langle \rho | n_u(\vec{R}+\vec{r}) n_d(\vec{R}) | \rho \rangle 
.
\end{eqnarray}
Since the spatial coordinate $\vec R$ is summed over, the correlators depend only on the spatial separation $\vec r$. 
The normal-ordered charge density operators are given by
\begin{eqnarray}
 n_u &=& :\bar{u} \gamma_4 u: \quad = \quad \bar{u} \gamma_4 u - \langle 0| \bar{u} \gamma_4 u |0\rangle
\\
 n_d &=& :\bar{d} \gamma_4 d: \quad = \quad \bar{d} \gamma_4 d - \langle 0| \bar{d} \gamma_4 d |0\rangle
.
\end{eqnarray}
Disconnected diagrams are excluded by the normal order.
As drawn in Fig.~\ref{fig}, the correlators are given by the products of four quark propagators.
When the temporal extent $T$ is large enough, they are dominated by the ground-state components.
Therefore, the correlators $C_\pi$ and $C_\rho$ are interpreted 
as the charge density distributions of the ground states of the pion and the $\rho$-meson.\footnote{These correlators are invariant under color SU(3) gauge transformation but not under electromagnetic U(1) gauge transformation. Thus the results should be interpreted as the charge density distributions with specific U(1) gauge fixing. This is equivalent to the gauge dependence of the wave function of a charged particle.}

To quantify the magnitude of the deformation, we define the mean-square interquark distance 
by the correlator as 
\begin{eqnarray}
 \bar{r}_{\pi,\rho}^2 &=& \bar{x}_{\pi,\rho}^2+\bar{y}_{\pi,\rho}^2+\bar{z}_{\pi,\rho}^2 \ = \ 2\bar{x}_{\pi,\rho}^2+\bar{z}_{\pi,\rho}^2
\\
 \bar{x}_{\pi,\rho}^2 &=& \frac{\sum_{\vec r} \ x^2 C_{\pi,\rho}(\vec{r})}{\sum_{\vec r} \ C_{\pi,\rho}(\vec{r})}
\\
 \bar{z}_{\pi,\rho}^2 &=& \frac{\sum_{\vec r} \ z^2 C_{\pi,\rho}(\vec{r})}{\sum_{\vec r} \ C_{\pi,\rho}(\vec{r})}
.
\end{eqnarray}
When the masses of $u$ and $d$ quarks are degenerate, the root-mean-square meson radius is estimated as $r_{\rm rms} = \bar{r}_{\pi,\rho}/2$.\footnote{This simple relation does not hold in a nonzero magnetic field.
The magnetic field induces the mass splitting between $u$ and $d$ quarks even if their bare masses are degenerate.}
The coefficient $1/2$ comes from the difference between a meson radius and an interquark distance. 

The formalism given in this section is applicable to other hadrons, even to neutral hadrons.
Although neutral hadrons seem insensitive to magnetic fields at a glance, 
they are affected through the coupling to the electric charges of quarks.
The inner structures of the neutral hadrons will be resolved by a strong enough magnetic field. 
The effects on the constituents will manifest themselves 
as the shift of spectra and the deformation of neutral hadrons. 
The calculation of these neutral mesons is, however, difficult because disconnected diagrams exist at nonzero magnetic fields \cite{Hidaka:2012mz}.
In this work, we study two of the simplest cases, the charged pion \eqref{eqpi} and the charged $\rho$-meson \eqref{eqrho}, which include only connected diagrams.

\section{Simulation results}

We used the same parameters as Ref.~\cite{Hidaka:2012mz}.
For gauge configurations, we used the quenched plaquette action without dynamical quarks with the coupling constant $\beta=5.9$.
The corresponding lattice spacing is $a\simeq 0.10$ fm.
For quark propagators, we used the unimproved Wilson fermion with the hopping parameter $\kappa = 0.1583$ for both $u$ and $d$ quarks.
The corresponding pion mass is $ m_\pi \simeq 0.49$ GeV.
Point sources are used for quark propagators.
The lattice size is $N_s^3 \times N_t = 16^3 \times 32$.
The minimum unit of $eB$ is $6\pi/N_s^2a^2 \simeq 0.27$ GeV$^2$.
Spatial boundary conditions are periodic.
Temporal boundary conditions are periodic for gluons and anti-periodic for quarks.
We checked the ground-state dominance of the correlators, as shown in an appendix.

\begin{figure}[t]
\begin{center}
\includegraphics[width=.49\textwidth]{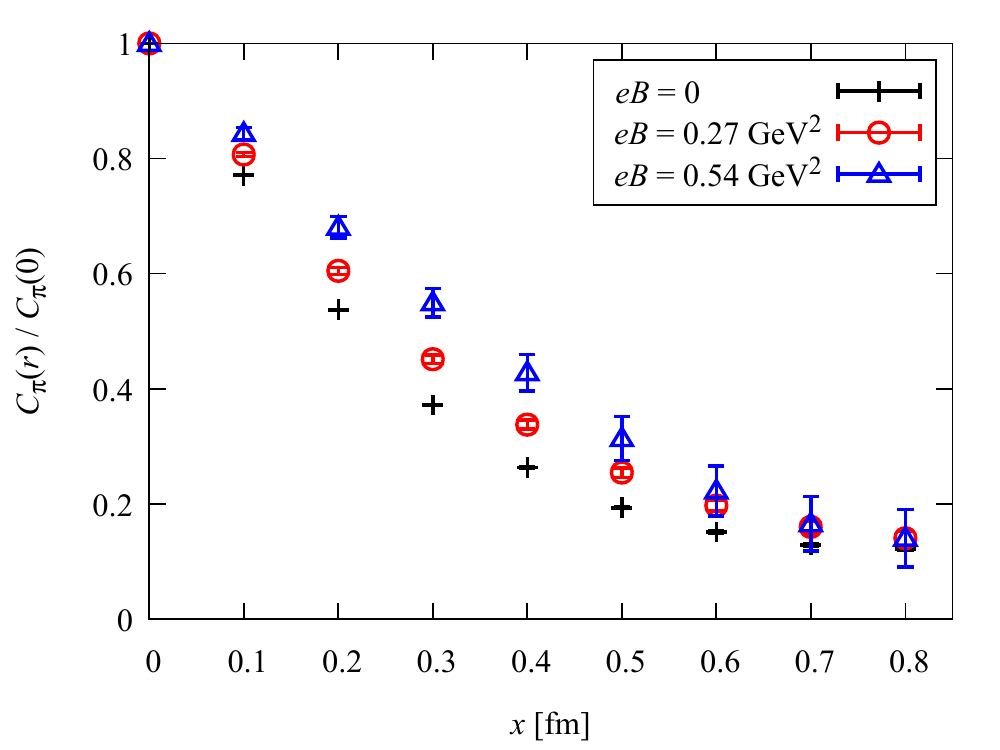}
\includegraphics[width=.49\textwidth]{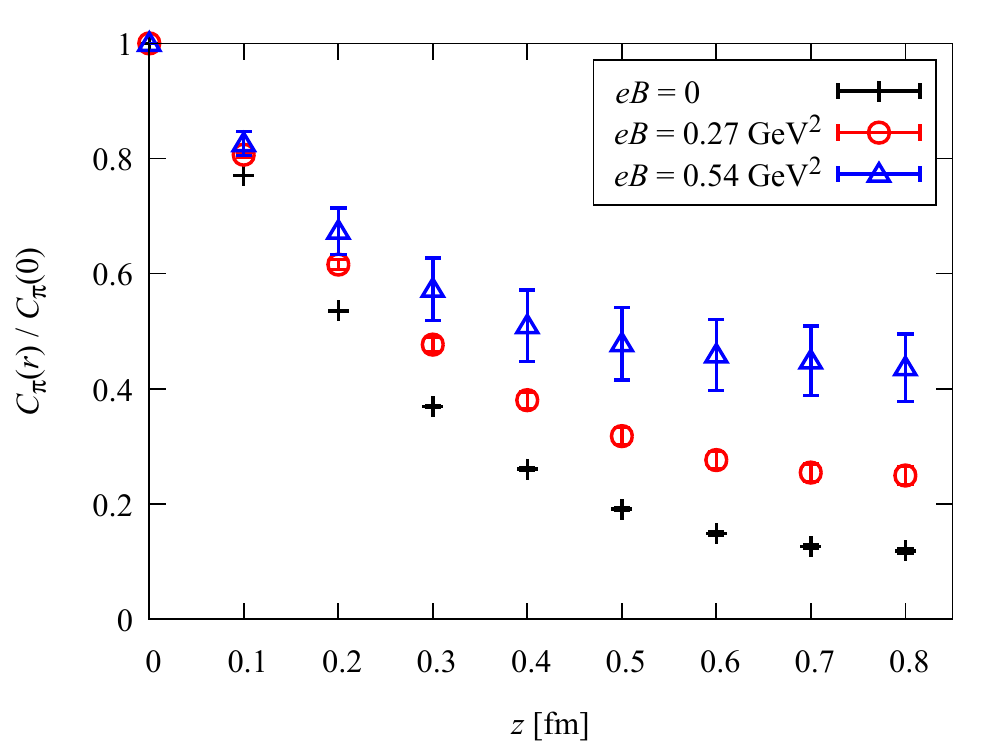}
\vspace{-0.25cm}
\caption{
\label{figXL}
The pion correlator $C_\pi$.
Left: the transverse direction to the magnetic field.
The other coordinates are fixed at $y=z=0$.
The behavior is the same in the $y$ direction.
Right: the longitudinal direction to the magnetic field.
The other coordinates are fixed at $x=y=0$.
}
\end{center}
\vspace{-0.25cm}
\end{figure}
\begin{figure}[h]
\begin{center}
\includegraphics[width=.49\textwidth]{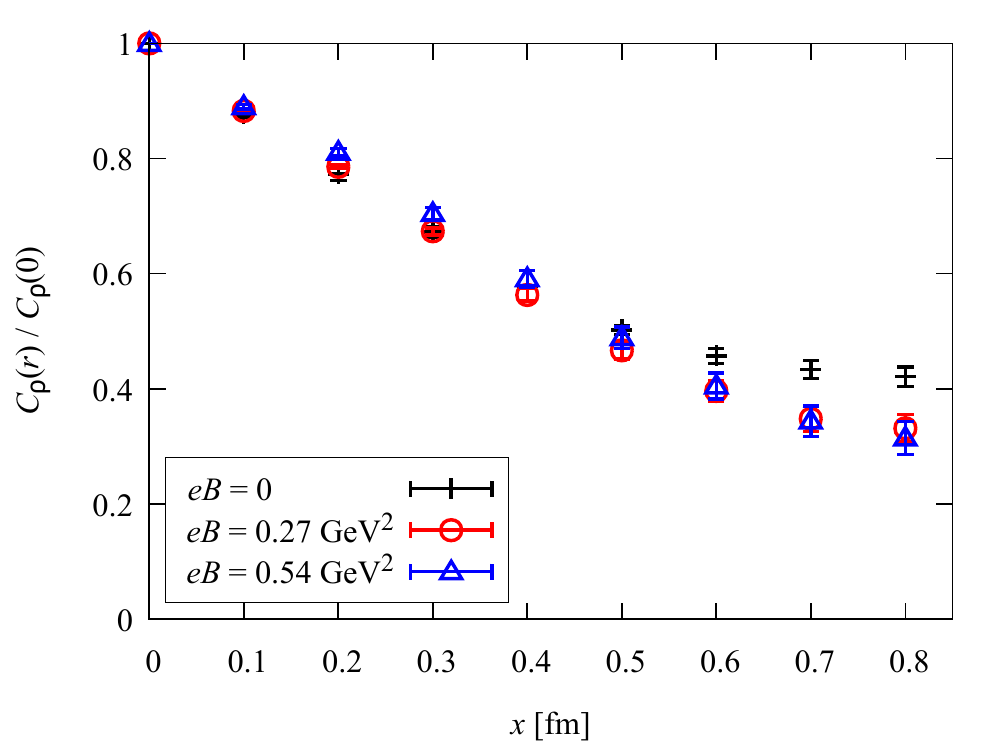}
\includegraphics[width=.49\textwidth]{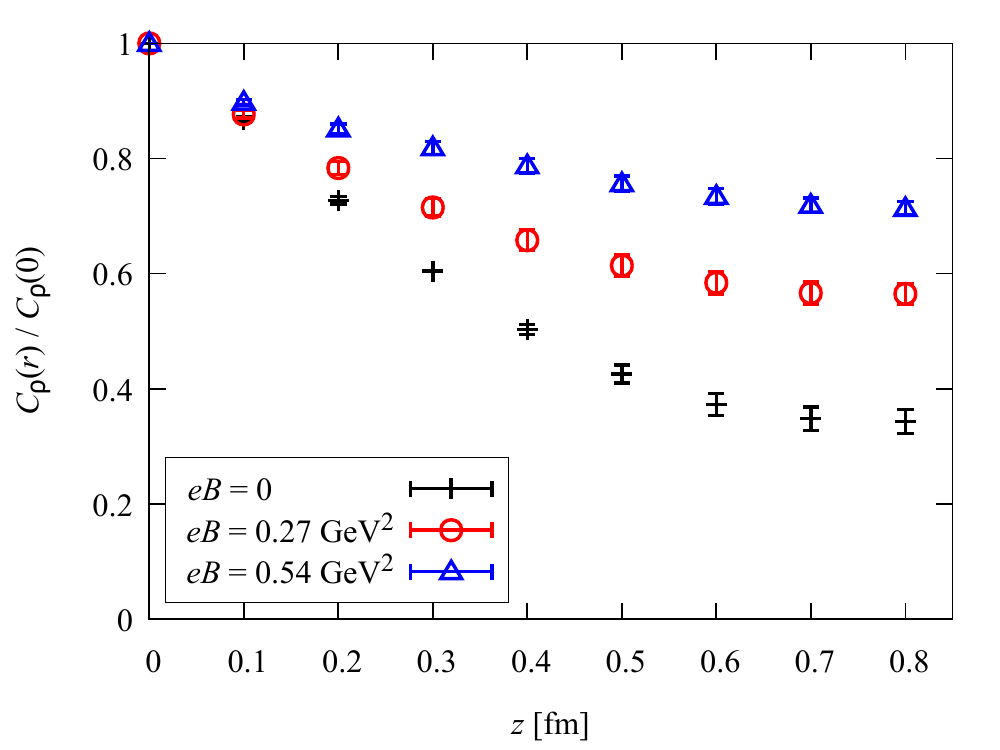}
\vspace{-0.25cm}
\caption{
\label{figXR}
The same as Fig.~\ref{figXL} but for the $\rho$-meson correlator $C_\rho$.
}
\end{center}
\vspace{-0.25cm}
\end{figure}

We first summarize the numerical results at $eB=0$. 
Shown in Figs.~\ref{figXL} and \ref{figXR} are the correlators 
for the pion and the $ \rho $-meson, respectively. 
The pion has a spherical shape. 
On the other hand, the $\rho$-meson has a slightly non-spherical shape 
with a larger transverse size than its longitudinal size. 
The mean-square values extracted from the correlators are summarized in Table \ref{tab}.
Our result corresponds to a pion radius $r_{\rm rms} \simeq 0.35$ fm.
This is smaller than the experimental value $r_{\rm exp} \simeq 0.67$ fm defined by a pion form factor \cite{PhysRevD.98.030001} but comparable to the previous lattice results in the same definition \cite{Alexandrou:2002nn,Alexandrou:2003qt}.

\begin{table}[t]
\begin{center}
\caption{
\label{tab}
The mean interquark distance $\bar{x}_{\pi,\rho}$, $\bar{z}_{\pi,\rho}$, and the meson mass $m_{\pi,\rho}$.
The numbers in parenthesis are statistical errors.
}
\begin{tabular}[t]{|c|l|l|l|l|l|l|}
\hline
$eB$ [GeV$^2$]& $\bar{x}_\pi$ [fm] & $\bar{z}_\pi$ [fm] & $\bar{x}_\rho$ [fm] & $\bar{z}_\rho$ [fm] & $m_\pi$ [GeV] & $m_\rho$ [GeV]
\\
\hline
0 & $0.400(1)$ & $0.400(1)$ & $0.437(5)$ & $0.425(7)$ & $0.485(4)$ & $0.822(28)$
\\
\hline
0.27 & $0.392(5)$ & $0.429(4)$ & $0.401(8)$ & $0.441(3)$ & $0.677(9)$ & $0.717(10)$
\\
\hline
0.54 & $0.359(29)$ & $0.450(4)$ & $0.399(6)$ & $0.453(1)$ & $0.927(34)$ & $0.664(4)$ 
\\
\hline
\end{tabular}
\end{center}
\end{table}

At $eB \ne 0$, the mesons are subject to the deformation caused by the magnetic field.
In Figs.~\ref{figXL} and \ref{figXR}, the longitudinal deformation of both the pion and the $\rho$-meson is clearly seen 
as elongated correlations in the $z$ direction. 
Transverse deformation is slight. 
The pion slightly enlarges and the $\rho$-meson slightly shrinks in the transverse direction 
(see also the discussion about the normalization effect in Sec.~\ref{sec:perspective}). 
From Table \ref{tab}, the deformation ratio is read off as 
$\bar z_\pi/\bar x_\pi \simeq 1.09$, $\bar z_\rho/\bar x_\rho \simeq 1.10$ at $eB \simeq 0.27$ GeV$^2$ and $\bar z_\pi/\bar x_\pi \simeq 1.25$, $\bar z_\rho/\bar x_\rho \simeq 1.14$ at $eB \simeq 0.54$ GeV$^2$. 
Thus, the mesons deform by about 10-20 \% at the strength of the magnetic field 
in the order of the pion mass square $ eB \sim m^2_\pi$. 
The pion mass in the current numerical setup is $m_\pi \simeq 0.49$ GeV
and is larger than the physical pion mass. 
Therefore, in the physical unit, these magnetic fields are stronger by an order than 
the realistic magnetic fields induced in relativistic heavy-ion collisions, of which the strengths 
are in the order of the physical pion mass square 
$eB  \sim m^2_\pi \sim 0.01$ GeV$^2$ (see Ref.~\cite{Hattori:2016emy} and references therein). 
However, given that hadron properties are governed by the pion mass, 
it might be possible that the mesons in heavy-ion collisions are deformed by 10-20 \%. 
The deformation will be milder in smaller magnetic fields but will be stronger for smaller quark masses.
The resultant deformation is determined by such a competition. 
To reach an unambiguous conclusion, we need realistic simulation with the physical pion mass.

\begin{figure}[t]
\begin{center}
\includegraphics[width=1\textwidth]{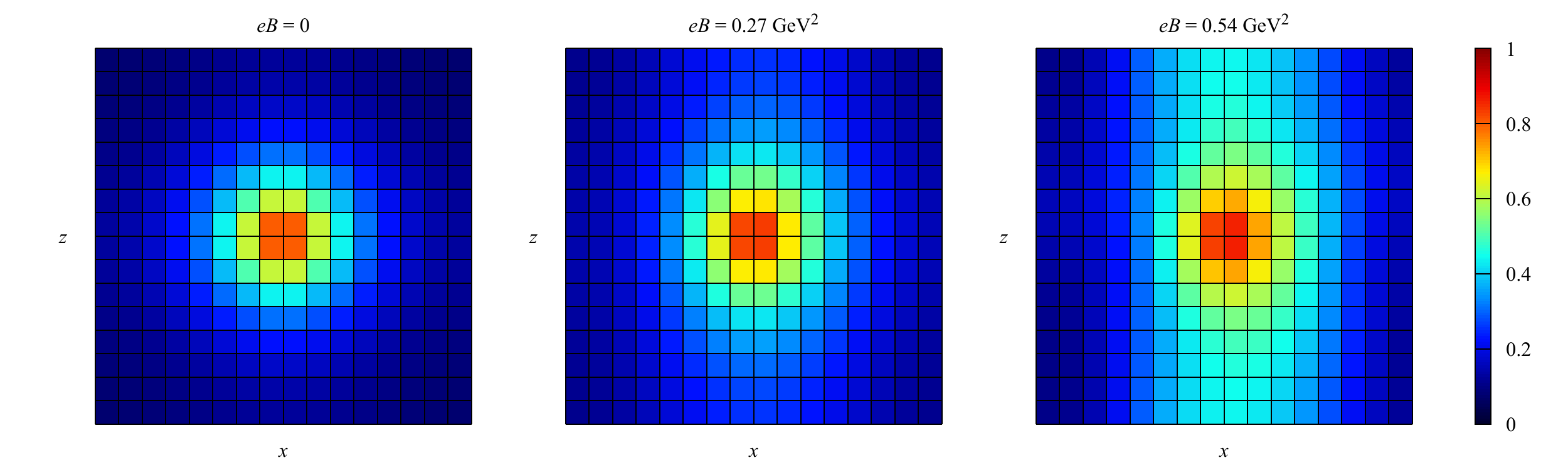}
\vspace{-0.5cm}
\caption{
\label{figXZL}
The pion correlator $C_\pi$ in the $x$-$z$ plane.
The normalized value $C_\pi(r)/C_\pi(0)$ is plotted by color gradation.
The other coordinate is fixed at $y=0$.
The magnetic field is applied to the $z$ direction.
Grids correspond to plaquettes.
}
\end{center}
\end{figure}
\begin{figure}[t]
\begin{center}
\includegraphics[width=1\textwidth]{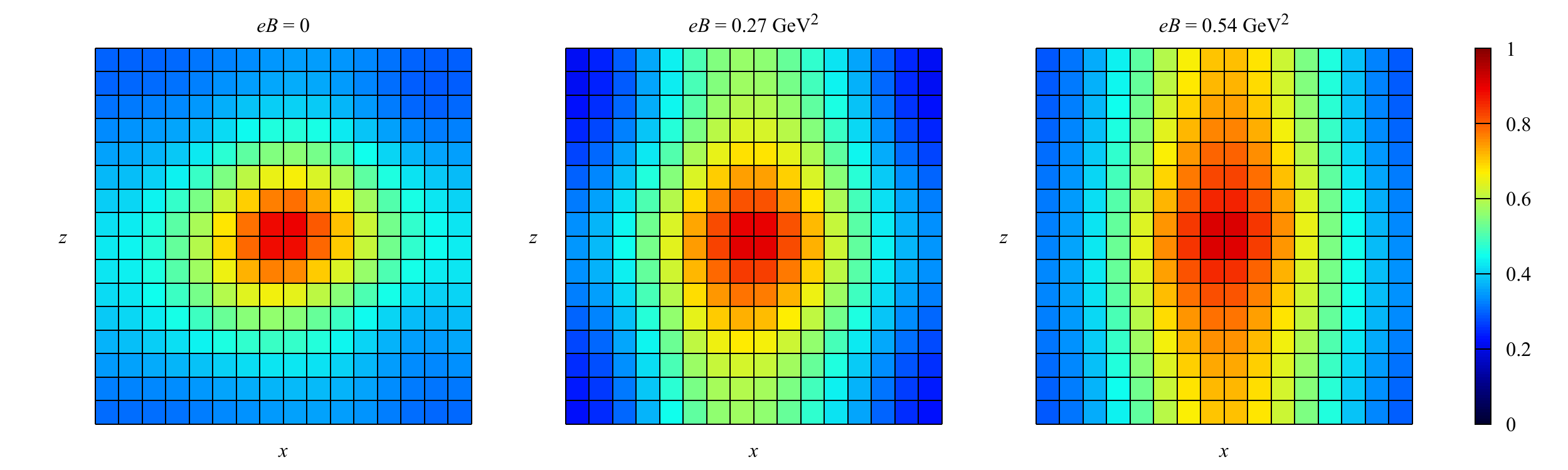}
\vspace{-0.5cm}
\caption{
\label{figXZR}
The same as Fig.~\ref{figXZL} but for the $\rho$-meson correlator $C_\rho$.
}
\end{center}
\end{figure}

The two-dimensional plots in Figs.~\ref{figXZL} and \ref{figXZR} 
show the deformation in the $x$-$z$ plane more intuitively. 
We clearly see the deformations of these mesons as we increase the magnetic field strength. 
The pion deforms from spherical to {\it prolate} (cigar-like) shape.
The $\rho$-meson deforms from {\it oblate} (disk-like) to {\it prolate} shape.
Note that the plotted distribution has a larger spatial extension than the actual size of each meson because the interquark distance is about twice as large as the meson radius.
The spatial extension seems to exceed the lattice size in the $z$-direction.
It is known that this density-density correlator is easily subject to finite size artifact, as pointed out in Ref.~\cite{Alexandrou:2002nn}.
We should use larger lattice sizes and take the infinite size limit for a more quantitative discussion.

\section{Discussions and perspective}

\label{sec:perspective}

In this paper, we have studied the deformation of mesons in magnetic fields 
on the basis of the density-density correlator in quenched lattice QCD. 
There are several theoretical scenarios for the deformation of meson shapes as raised below. 
Since they are not completely independent mechanisms, 
we may not clearly distinguish them in a rigorous sense of QCD.
Nevertheless, it would be valuable to interpret the simulation results from phenomenological viewpoints.

\begin{itemize}

\item{\it Mass shift.}
Meson masses depend on the magnetic field strength.
As shown in Table \ref{tab}. the pion mass is increased and 
the polarized $\rho$-meson mass is decreased by the magnetic fields (see also Ref.~\cite{Hidaka:2012mz}).
The mass shift can be regarded as the change of binding strength. 
Thus, the pion and the $\rho$-meson tend to enlarge and shrink in all directions, respectively. 
This effect is consistent with the slight transverse deformation seen in Figs.~\ref{figXL} and \ref{figXR}.

\item{\it Normalization.}
When the volume integral of a wave function is normalized to unity, 
the wave function elongates in one direction as it shrinks in other directions, and vice versa. 
The mean-square values shown in Table \ref{tab} are affected by this normalization effect. 
On the other hand, the correlators in all the figures are normalized to unity at the origin, as $ C_{\pi,\rho}(r) / C_{\pi,\rho}(0) $,
so that the deformation seen in those figures is nothing to do with this effect. 
This difference may be responsible for the two opposite observations 
that the pion correlator in Fig.~\ref{figXL} elongates in the transverse direction whereas the mean-square value $ \bar x_\pi $ decreases. 
The normalization effect could have such a big impact on $ \bar x_\pi $ because 
the large elongation in the longitudinal direction 
has to be compensated with the same magnitude of the shrinkage in the transverse direction
which overwhelms the slight transverse elongation of the correlator itself. 

\item{\it Cyclotron motion and anisotropic fluctuations.}
The transverse motions of charged particles in a magnetic field are confined in cyclotron orbits.
They shrink as the magnetic field strength increases.
Therefore, the motions of quarks become anisotropic  
and oriented along the longitudinal direction. 
As a consequence, the longitudinal fluctuation in the coordinate space is strongly enhanced
as compared to the transverse one. 
This effect is clear from the longitudinal elongation of both mesons. 

\item{\it Landau quantization.}
As the cyclotron orbit shrinks, quantum effects become prominent. 
Indeed, the harmonic motion in the transverse plane leads to the Landau quantization. 
This quantum effect has not been clearly identified in our simulation,
but might manifest itself in much stronger magnetic fields.
In the limit of strong magnetic fields, the quarks in the ground states, called the lowest Landau levels, 
play dominant roles (cf., Ref.~\cite{Hattori:2015aki} for an analytical study). 
The transverse radius will decrease
and take the smallest value for the lowest Landau levels, as computed with the corresponding wave functions.\footnote{
While the wave function depends on the gauge of an external magnetic field, 
the expectation value of the radius from the center of cyclotron motion is a gauge-invariant quantity. 
In the momentum space, the transverse fluctuation grows with the magnetic field strength as expected. 
} 

\item{\it Rotational modes in charged mesons.}
The interquark attractive force  plays a counterpart of an ``electric field'' 
giving rise to the drift velocity $ v_{\rm d} = \sigma/(qB)  $ with a string tension $ \sigma$. 
The quark and antiquark drift in the opposite directions, 
so that the drift motion induces internal rotational modes of charged mesons (see Fig.~5 in Ref.~\cite{Hattori:2015aki}). 
The radius of the rotational mode may be estimated from the balance between 
the centrifugal force and the string tension $ \sigma \sim m v_d^2/r  $, suggesting that $ r \sim m \sigma/(qB)^2 $. 
Thus, the transverse radius decreases as the centrifugal barrier 
diminishes with a smaller drift velocity when the magnetic field strength is increased. 

\item{\it Anisotropic interquark interaction.}
The anisotropic sea-quark excitations result in an anisotropic strength of interquark interaction,
as shown by lattice QCD simulation \cite{Bonati:2014ksa,Bonati:2016kxj,Bonati:2017uvz,Bonati:2018uwh}. 
Perturbative computation also indicates an anisotropic screening effect 
in a strong magnetic field~\cite{Fukushima:2011nu, Hattori:2012je, Hattori:2017xoo}. 
In our results, this effect will be small because almost no sea-quark effect 
is taken into account in quenched QCD, where loop diagrams are neglected and only hairpin diagrams exist. 

\end{itemize}

In future work, we can apply the same analysis method to other mesons and baryons.
Needless to say, the application to nucleons is the most important from a phenomenological viewpoint.
It will be relevant for protons in relativistic heavy-ion collisions and neutrons in neutron stars. 
It will be also interesting to investigate $ \Lambda $ hyperons for an application to relativistic heavy-ion collisions. 
Recent experimental data show that there is a splitting between 
the magnitudes of the spin polarization of $  \Lambda$ and $ \bar \Lambda $~\cite{STAR:2017ckg, Adam:2018ivw}, 
which could imply a spin polarization along the magnetic field on top of the charge-blind polarization effect 
in response to the local vorticity and/or the global rotation of the quark-gluon plasma. 

In general, the definition of the spatial distribution of hadrons is not unique.
We can adopt other definitions instead of the density-density correlator.
Measuring other observables, such as a form factor, allows us a more direct comparison to experimental observables. 
For a more quantitative discussion, we should refine the simulation, i.e., work in full QCD, 
use the physical quark mass, and take the continuum and infinite-volume limits.

\ack
A.~Y.~was supported by JSPS KAKENHI Grant Number 15K17624. 
The numerical simulations were carried out on SX-ACE in Osaka University.

\appendix

\section{Checking the ground-state dominance}

\begin{figure}[h]
\begin{center}
 \includegraphics[width=.6\textwidth]{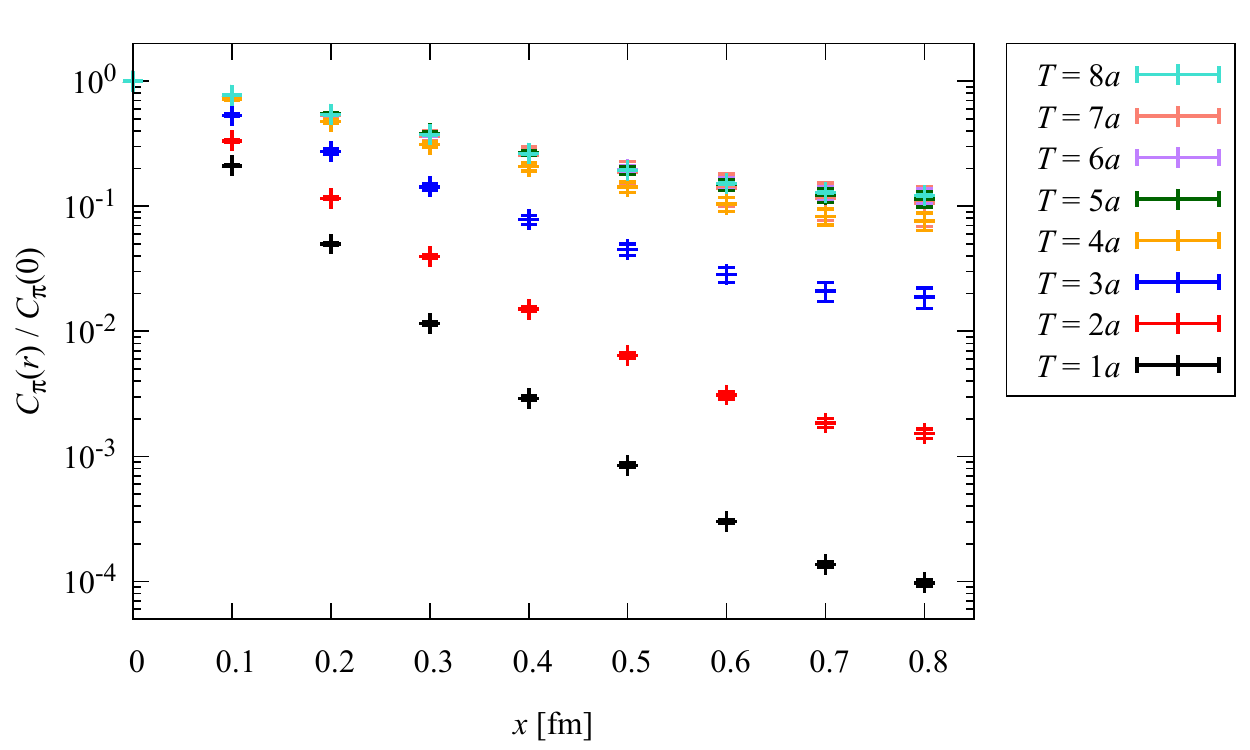}
\caption{
\label{figTL}
The $T$-dependence of the pion correlator $C_\pi$ at $eB=0$.
}
\end{center}
\end{figure}

In this appendix, we check the ground-state dominance of the correlators.
The correlator in Fig.~\ref{fig} includes not only a ground state but also excited states.
For the correlator to be dominated by the ground state, the temporal extent $T$ must be large enough.
One typical example of the $T$-dependence of the correlator is shown in Fig.~\ref{figTL}.
The correlator depends on $T$ in $T \le 5a$, but it is independent of $T$ in $T \ge 5a$.
This means that the ground-state dominance is achieved by taking $T \ge 5a$.
The maximal temporal extent is $T = N_ta/4$ because of (anti-)periodicity.
In this work, we took the maximal temporal extent $T = N_ta/4=8a$ to calculate the correlators.

\bibliographystyle{ptephy}
\bibliography{paper}

\begin{thebibliography}{10}

\bibitem{Harding:2006qn}
A.~K. Harding and D.~Lai, Rept. Prog. Phys., {\bf 69}, 2631 (2006),
  {{arXiv:astro-ph/0606674}}.

\bibitem{Hattori:2016emy}
K.~Hattori and X.-G. Huang, Nucl. Sci. Tech., {\bf 28}, 26 (2017),
  {{arXiv:1609.00747}}.

\bibitem{Buividovich:2009wi}
P.~V. Buividovich, M.~N. Chernodub, E.~V. Luschevskaya, and M.~I. Polikarpov,
  Phys. Rev. D, {\bf 80}, 054503 (2009),  {{arXiv:0907.0494}}.

\bibitem{Yamamoto:2011gk}
A.~Yamamoto, Phys. Rev. Lett., {\bf 107}, 031601 (2011),  {{arXiv:1105.0385}}.

\bibitem{Yamamoto:2011ks}
A.~Yamamoto, Phys. Rev. D, {\bf 84}, 114504 (2011),  {{arXiv:1111.4681}}.

\bibitem{Bali:2014vja}
G.~S. Bali, F.~Bruckmann, G.~Endrodi, Z.~Fodor, S.~D. Katz, and A.~Schafer,
  JHEP, {\bf 04}, 129 (2014),  {{arXiv:1401.4141}}.

\bibitem{Fukushima:2015wck}
K.~Fukushima, K.~Hattori, H.-U. Yee, and Y.~Yin, Phys. Rev. D, {\bf 93}, 074028
  (2016),  {{arXiv:1512.03689}}.

\bibitem{Hattori:2016cnt}
K.~Hattori and D.~Satow, Phys. Rev. D, {\bf 94}, 114032 (2016),
  {{arXiv:1610.06818}}.

\bibitem{Hattori:2016lqx}
K.~Hattori, S.~Li, D.~Satow, and H.-U. Yee, Phys. Rev. D, {\bf 95}, 076008
  (2017),  {{arXiv:1610.06839}}.

\bibitem{Fukushima:2017lvb}
K.~Fukushima and Y.~Hidaka, Phys. Rev. Lett., {\bf 120}, 162301 (2018),
  {{arXiv:1711.01472}}.

\bibitem{Li:2017tgi}
S.~Li and H.-U. Yee, Phys. Rev. D, {\bf 97}, 056024 (2018),
  {{arXiv:1707.00795}}.

\bibitem{Hattori:2017qih}
K.~Hattori, X.-G. Huang, D.~H. Rischke, and D.~Satow, Phys. Rev. D, {\bf 96},
  094009 (2017),  {{arXiv:1708.00515}}.

\bibitem{Bali:2012zg}
G.~S. Bali, F.~Bruckmann, G.~Endrodi, Z.~Fodor, S.~D. Katz, and A.~Schafer,
  Phys. Rev. D, {\bf 86}, 071502 (2012),  {{arXiv:1206.4205}}.

\bibitem{Bruckmann:2013oba}
F.~Bruckmann, G.~Endrodi, and T.~G. Kovacs, JHEP, {\bf 04}, 112 (2013),
  {{arXiv:1303.3972}}.

\bibitem{DElia:2011koc}
M.~D'Elia and F.~Negro, Phys. Rev. D, {\bf 83}, 114028 (2011),
  {{arXiv:1103.2080}}.

\bibitem{Ilgenfritz:2013ara}
E.~M. Ilgenfritz, M.~Muller-Preussker, B.~Petersson, and A.~Schreiber, Phys.
  Rev. D, {\bf 89}, 054512 (2014),  {{arXiv:1310.7876}}.

\bibitem{DElia:2018xwo}
M.~D'Elia, F.~Manigrasso, F.~Negro, and F.~Sanfilippo, Phys. Rev. D, {\bf 98},
  054509 (2018),  {{arXiv:1808.07008}}.

\bibitem{Gusynin:1994xp}
V.~P. Gusynin, V.~A. Miransky, and I.~A. Shovkovy, Phys. Lett. B, {\bf 349},
  477 (1995),  {{arXiv:hep-ph/9412257}}.

\bibitem{Gusynin:1997kj}
V.~P. Gusynin and I.~A. Shovkovy, Phys. Rev. D, {\bf 56}, 5251 (1997),
  {{arXiv:hep-ph/9704394}}.

\bibitem{Lee:1997uh}
D.~S. Lee, C.~N. Leung, and Y.~J. Ng, Phys. Rev. D, {\bf 57}, 5224 (1998),
  {{arXiv:hep-th/9711126}}.

\bibitem{Fukushima:2012xw}
K.~Fukushima and J.~M. Pawlowski, Phys. Rev. D, {\bf 86}, 076013 (2012),
  {{arXiv:1203.4330}}.

\bibitem{Lee:2008qf}
F.~X. Lee, S.~Moerschbacher, and W.~Wilcox, Phys. Rev. D, {\bf 78}, 094502
  (2008),  {{arXiv:0807.4150}}.

\bibitem{Bali:2011qj}
G.~S. Bali, F.~Bruckmann, G.~Endrodi, Z.~Fodor, S.~D. Katz, S.~Krieg,
  A.~Schafer, and K.~K. Szabo, JHEP, {\bf 02}, 044 (2012),
  {{arXiv:1111.4956}}.

\bibitem{Hidaka:2012mz}
Y.~Hidaka and A.~Yamamoto, Phys. Rev. D, {\bf 87}, 094502 (2013),
  {{arXiv:1209.0007}}.

\bibitem{Primer:2013pva}
T.~Primer, W.~Kamleh, D.~Leinweber, and M.~Burkardt, Phys. Rev. D, {\bf 89},
  034508 (2014),  {{arXiv:1307.1509}}.

\bibitem{Beane:2014ora}
S.~R. Beane, E.~Chang, S.~Cohen, W.~Detmold, H.~W. Lin, K.~Orginos, A.~Parreno,
  M.~J. Savage, and B.~C. Tiburzi, Phys. Rev. Lett., {\bf 113}, 252001 (2014),
  {{arXiv:1409.3556}}.

\bibitem{Chang:2015qxa}
E.~Chang, W.~Detmold, K.~Orginos, A.~Parreno, M.~J. Savage, B.~C. Tiburzi, and
  S.~R. Beane, Phys. Rev. D, {\bf 92}, 114502 (2015),  {{arXiv:1506.05518}}.

\bibitem{Agadjanov:2016cjc}
A.~Agadjanov, U.-G. Meissner, and A.~Rusetsky, Phys. Rev. D, {\bf 95}, 031502
  (2017),  {{arXiv:1610.05545}}.

\bibitem{Bali:2017ian}
G.~S. Bali, B.~B. Brandt, G.~Endrodi, and B.~Glassle, Phys. Rev. D, {\bf 97},
  034505 (2018),  {{arXiv:1707.05600}}.

\bibitem{Bali:2018sey}
G.~S. Bali, B.~B. Brandt, G.~Endrodi, and B.~Glassle, Phys. Rev. Lett., {\bf
  121}, 072001 (2018),  {{arXiv:1805.10971}}.

\bibitem{Burkardt:1994pw}
M.~Burkardt, J.~M. Grandy, and J.~W. Negele, Annals Phys., {\bf 238}, 441
  (1995),  {{arXiv:hep-lat/9406009}}.

\bibitem{Alexandrou:2002nn}
C.~Alexandrou, P.~de~Forcrand, and A.~Tsapalis, Phys. Rev. D, {\bf 66}, 094503
  (2002),  {{arXiv:hep-lat/0206026}}.

\bibitem{Alexandrou:2003qt}
C.~Alexandrou, P.~de~Forcrand, and A.~Tsapalis, Phys. Rev. D, {\bf 68}, 074504
  (2003),  {{arXiv:hep-lat/0307009}}.

\bibitem{Endrodi:2013cs}
G.~Endrodi, JHEP, {\bf 04}, 023 (2013),  {{arXiv:1301.1307}}.

\bibitem{Hattori:2015aki}
K.~Hattori, T.~Kojo, and N.~Su, Nucl. Phys. A, {\bf 951}, 1 (2016),
  {{arXiv:1512.07361}}.

\bibitem{Fukushima:2016vix}
K.~Fukushima and Y.~Hidaka, Phys. Rev. Lett., {\bf 117}, 102301 (2016),
  {{arXiv:1605.01912}}.

\bibitem{Fukushima:2012kc}
K.~Fukushima and Y.~Hidaka, Phys. Rev. Lett., {\bf 110}, 031601 (2013),
  {{arXiv:1209.1319}}.

\bibitem{Kojo:2012js}
T.~Kojo and N.~Su, Phys. Lett. B, {\bf 720}, 192 (2013),  {{arXiv:1211.7318}}.

\bibitem{Alford:2013jva}
J.~Alford and M.~Strickland, Phys. Rev. D, {\bf 88}, 105017 (2013),
  {{arXiv:1309.3003}}.

\bibitem{Suzuki:2016kcs}
K.~Suzuki and T.~Yoshida, Phys. Rev. D, {\bf 93}, 051502 (2016),
  {{arXiv:1601.02178}}.

\bibitem{Yoshida:2016xgm}
T.~Yoshida and K.~Suzuki, Phys. Rev. D, {\bf 94}, 074043 (2016),
  {{arXiv:1607.04935}}.

\bibitem{AlHashimi:2008hr}
M.~H. Al-Hashimi and U.~J. Wiese, Annals Phys., {\bf 324}, 343 (2009),
  {{arXiv:0807.0630}}.

\bibitem{PhysRevD.98.030001}
M.~Tanabashi {\it et al.}, Phys. Rev. D, {\bf 98}, 030001 (2018).

\bibitem{Bonati:2014ksa}
C.~Bonati, M.~D'Elia, M.~Mariti, M.~Mesiti, F.~Negro, and F.~Sanfilippo, Phys.
  Rev. D, {\bf 89}, 114502 (2014),  {{arXiv:1403.6094}}.

\bibitem{Bonati:2016kxj}
C.~Bonati, M.~D'Elia, M.~Mariti, M.~Mesiti, F.~Negro, A.~Rucci, and
  F.~Sanfilippo, Phys. Rev. D, {\bf 94}, 094007 (2016),  {{arXiv:1607.08160}}.

\bibitem{Bonati:2017uvz}
C.~Bonati, M.~D'Elia, M.~Mariti, M.~Mesiti, F.~Negro, A.~Rucci, and
  F.~Sanfilippo, Phys. Rev. D, {\bf 95}, 074515 (2017),  {{arXiv:1703.00842}}.

\bibitem{Bonati:2018uwh}
C.~Bonati, S.~Cali, M.~D'Elia, M.~Mesiti, F.~Negro, A.~Rucci, and
  F.~Sanfilippo, Phys. Rev. D, {\bf 98}, 054501 (2018),  {{arXiv:1807.01673}}.

\bibitem{Fukushima:2011nu}
K.~Fukushima, Phys. Rev. D, {\bf 83}, 111501 (2011),  {{arXiv:1103.4430}}.

\bibitem{Hattori:2012je}
K.~Hattori and K.~Itakura, Annals Phys., {\bf 330}, 23 (2013),
  {{arXiv:1209.2663}}.

\bibitem{Hattori:2017xoo}
K.~Hattori and D.~Satow, Phys. Rev. D, {\bf 97}, 014023 (2018),
  {{arXiv:1704.03191}}.

\bibitem{STAR:2017ckg}
L.~Adamczyk et~al., Nature, {\bf 548}, 62 (2017),  {{arXiv:1701.06657}}.

\bibitem{Adam:2018ivw}
J.~Adam et~al., Phys. Rev. C, {\bf 98}, 014910 (2018),  {{arXiv:1805.04400}}.

\end{thebibliography}

\end{document}